\documentclass[a4paper,11pt]{article}

\usepackage{amsmath}
\usepackage{amssymb}
\usepackage{color}
\usepackage[dvips]{graphicx}
\usepackage[dvips]{psfrag}
\usepackage{cite}

\makeatletter
\@addtoreset{equation}{section}
\renewcommand{\theequation}{\thesection.\@arabic\c@equation}
\makeatother

\makeatletter
\renewcommand\appendix{\par
  \setcounter{section}{0}%
  \setcounter{subsection}{0}%
  \gdef\thesection{Appendix \@Alph\c@section }
  \renewcommand{\theequation}
  {\Alph{section}.\arabic{equation}}
}
\makeatother

\def \be {\begin{equation}}
\def \ee {\end{equation}}
\def \ba {\begin{array}}
\def \ea {\end{array}}
\def \bea{\begin{eqnarray}}
\def \eea{\end{eqnarray}}

\def \a {\alpha}
\def \b {\beta}
\def \g {\gamma}
\def \G {\Gamma}
\def \d {\delta}
\def \D {\Delta}
\def \e {\epsilon}

\def \m {\mu}
\def \n {\nu}

\def \l {\lambda}
\def \L {\Lambda}
\def \s {\sigma}
\def \S {\Sigma}
\def \r {\rho}
\def \o {\omega}
\def \O {\Omega}
\def \th {\theta}

\def \t {\tau}

\def \p {\partial}
\def \f {\frac}
\def \na {\nabla}

\def \nn {\nonumber}
\def \ma {\mathcal}

\def \lt {\left}
\def \rt {\right}

\def \ra {\rightarrow}
\def \sr {\sqrt}
\def \td {\tilde}
\def \hs {\hspace}
\def \pp {\propto}
\def \inf {\infty}

\title{\textbf{Quasinormal Modes and Hidden Conformal Symmetry of Warped dS$_3$ Black Hole}}
\author{
Bin Chen$^{1,2,3}$\footnote{bchen01@pku.edu.cn}
,
Jia-ju Zhang$^1$\footnote{jjzhang@pku.edu.cn}
 and
Jian-dong Zhang$^1$\footnote{mksws@pku.edu.cn}
}
\date{}

\begin{document}

\maketitle

\begin{center}
\vspace{5mm}
{\it
$^{1}$Department of Physics, and State Key Laboratory of Nuclear Physics and Technology,\\
Peking University, Beijing 100871, P.R. China\\
\vspace{2mm}
$^{2}$Center for High Energy Physics, Peking University, Beijing 100871, P.R. China\\
$^3$Kavli Institute for Theoretical Physics China,
CAS, Beijing 100190, P.R. China
}
\vspace{10mm}
\end{center}

\begin{abstract}

In this paper, we analytically calculate the quasinormal modes of scalar, vector, tensor, and spinor perturbations of the warped dS$_3$ black hole. There are two horizons for the warped dS$_3$ black hole, namely, the black hole horizon $r_b$ and the cosmological horizon $r_c$. In the calculation, we impose the ingoing boundary condition at the black hole horizon and the outgoing boundary condition at the cosmological horizon. We also investigate the hidden conformal symmetry of the warped dS$_3$ black hole in the region between the black hole horizon and the cosmological horizon $r_b<r<r_c$. We use the hidden conformal symmetry to construct the quasinormal modes in an algebraic way and find that the results agree with the analytically ones. It turns out that the frequencies of the quasinormal modes could be identified with the poles in the thermal boundary-boundary correlators.

\end{abstract}

\baselineskip 18pt

\thispagestyle{empty}

\newpage

\section{Introduction}

The study of quasinormal modes is important in understanding black hole physics \cite{QNMs}. The quasinormal modes are defined as the perturbations subject to proper boundary conditions at the horizons and the infinity. The frequencies of the quasinormal modes have negative imaginary parts, indicating  the perturbations undergo damped oscillations. In the AdS/CFT correspondence, the quasinormal modes of black holes correspond to the operators perturbing the thermal equilibrium in dual finite temperature conformal field theory (CFT). The frequencies of the quasinormal modes agree with the poles of retarded Green's function of the dual CFT \cite{Horowitz:1999jd}. The scalar perturbation under Ba{\~n}ados-Teitelboim-Zanelli (BTZ) black hole background was considered in \cite{Chan:1996yk}. In the remarkable paper \cite{Birmingham:2001pj}, the quantitative agreement has been confirmed for the scalar, vector and spinor perturbations of three-dimensional BTZ black holes.

Inspired by the study of AdS$_3$/CFT$_2$ correspondence in
the BTZ black hole, a new kind of AdS/CFT correspondence was proposed in \cite{Anninos:2008fx} for warped spacetimes. It was pointed out that for the spacelike stretched and the null warped AdS$_3$ black holes, which are the classical solutions of three-dimensional topological massive gravity with a negative cosmological constant, there exist dual two-dimensional conformal field theory descriptions. The central charges of dual CFT have been confirmed in \cite{Compere:2008cv, Blagojevic:2009ek} for spacelike case.
The quasinormal modes of  scalar, vector and spinor perturbations of the spacelike stretched and the null warped AdS$_3$ black holes were calculated in \cite{Chen:2009rf,Chen:2009hg}. Moreover, the retarded Green's functions for scalar and vector fields in the spacelike stretched and the null warped AdS$_3$ black holes were obtained in \cite{Chen:2009cg}. It was found that the real-time correlators from the computations in the gravity side are well consistent with the CFT predictions, supporting the warped AdS/CFT correspondence. The similar study has been applied to self-dual warped AdS$_3$ black holes\cite{Chen:2010qm,Li:2010sv}.

The quasinormal modes in three-dimensional black holes could be constructed with the help of the hidden conformal symmetry.
The hidden conformal symmetry of black holes was first investigated in the low frequency scattering off generic nonextreme  Kerr black hole \cite{Castro:2010fd}. The symmetry is {\it hidden} in the sense that it appears in the solution space of the wave equation rather than the isometry of the geometry. More precisely, it was found that the scalar equation of motion in the near region could be written in terms of SL(2,R) Casimir.  In particular, for three-dimensional black holes, the hidden conformal symmetry could often be defined in the whole spacetime rather than just the near region in a higher-dimensional case. This makes the study of the hidden conformal symmetry in three-dimensional black holes more interesting and tractable. In the paper \cite{Chen:2010ik}, it was shown that
for general tensor fields, their equations of motion in three-dimensional BTZ and warped black holes could be written in terms of Lie-induced SL(2,R) Casimir as well. This fact indicates that the hidden conformal symmetry is an intrinsic property of the black hole. Furthermore in \cite{Chen:2010ik} the quasinormal modes of scalar, vector and tensor perturbations of various kinds of three-dimensional black holes have been constructed in an algebraic way using the ladder generators of SL(2,R) algebra.  The similar construction was then generalized to extremal three-dimensional black holes in \cite{Chen:2010sn} using the hidden conformal symmetry of extremal black holes proposed in \cite{Chen:2010fr}.

It would be interesting to generalize the above investigations to the de Sitter spacetime. It has been conjectured that there is also a holographic CFT description of quantum gravity in de Sitter spacetime, similar to AdS/CFT correspondence. For a de Sitter spacetime, there exist a cosmological horizon, whose properties are much similar to a black hole event horizon, carrying Hawking temperature and whose area gives the entropy of the de Sitter spacetime \cite{Gibbons:1977mu}.
The three-dimensional (Kerr-)dS/CFT was studied in \cite{Park} using the Chern-Simons theory with boundaries.
It was conjectured that there is dual Euclidean CFT living at the future timelike infinity $\ma I^+$ \cite{Strominger:2001pn,{Bousso:2001mw}}. In general the dual CFT may be nonunitary and the operators in it may contain complex conformal weights. For the recent studies in this direction, see \cite{Anninos:2010zf,Anninos:2011jp,Anninos:2011af}. Moreover, the holographic description of four-dimensional Kerr-dS black hole was proposed in \cite{Anninos:2009yc,Anninos:2010gh}. The analysis there was similar to that in \cite{Hartman:2008pb}, except that it was done at the cosmological horizon instead of the black hole horizon. On the other hand, the warped AdS/CFT correspondence has also been generalized to the de Sitter case. From the 3D TMG with a positive cosmological constant, there exist warped de Sitter solutions as well \cite{Anninos:2009jt}. Even for the pure dS$_3$ solution, the partition functions on the lens space turns out to have better behavior \cite{Castro:2011xb,Castro:2011ke}. Very recently, the holography for the warped dS$_3$ spacetimes were investigated in \cite{Anninos:2011vd}. In the paper, the authors studied the thermodynamics and the asymptotic structure of a family of warped dS$_3$ geometries, and then they conjectured that quantum gravity in asymptotically warped dS$_3$ is holographically dual to a two-dimensional conformal field theory living at $\ma I^+$ with the central charges $(c_L,c_R)$.

In this paper, we study several aspects of the warped dS/CFT correspondence. We solve analytically the equations of motion of the scalar, vector, tensor and spinor perturbations around the warped dS$_3$ black hole. In particular, we analyze the scalar solutions both in the physical region between the black hole horizon and the cosmological horizon and the unphysical region beyond the cosmological horizon. Moreover, we compute the boundary-boundary correlators in thermal background. Furthermore, we obtain the quasinormal modes of various perturbations in the warped dS$_3$ black hole, by imposing the ingoing boundary condition at the black hole horizon and the outgoing boundary condition at the cosmological horizon.  We also investigate the hidden conformal symmetry of warped dS$_3$ black hole in the physical region.  Finally, we construct the quasinormal modes of scalar, vector and tensor perturbations using the hidden conformal symmetry. It turns out that the frequencies of the quasinormal modes could be identified with the poles in the thermal boundary-boundary correlators.

The remaining parts of the paper are organized as follows. In Sec. 2 we give a brief review of the warped dS$_3$ black hole. Especially, we discuss the entropy of the black hole horizon. In Sec. 3 we solve the scalar equation in the warped dS$_3$ black hole background. From the late-time behavior, we read the thermal boundary-boundary correlators. In Sect. 4 we get the quasinormal modes of various perturbations by imposing proper boundary conditions. In Sec. 5 we investigate the hidden conformal symmetry of the warped dS$_3$ black hole and construct the quasinormal modes in an algebraic way. We end with some discussions in Sec. 6.

\section{Warped dS$_3$ Black Hole}

The contents of this section are mainly based on  \cite{Anninos:2011vd}, and the readers can see more details therein. Besides reviewing the thermodynamics of the warped black hole at the cosmological horizon, we also give a brief discussion of the thermodynamics at the black hole horizon.

The action for three-dimensional topological massive gravity (TMG) with a positive cosmological constant $\L=1/\ell^2$ is
\be
I_{TMG}=\f{1}{16\pi G}\int d^3x \sr{-g} \lt[ R-\f{2}{\ell^2}  +\f{1}{2\m}\e^{\l\m\n}\G^\r_{\l\s}
                             \lt( \p_\m\G^\s_{\r\n} +\f{2}{3}\G^\s_{\m\t}\G^\t_{\n\r} \rt) \rt].
\ee
Besides the usual Einstein-Hilbert term, there is also the so called Chern-Simons term in the action. The corresponding equation of motion is
\be
R_{\m\n}-\f{1}{2}R g_{\m\n}+\f{1}{\ell^2}g_{\m\n}+\f{1}{\m}C_{\m\n}=0,
\ee
and here we have the traceless symmetric Cotton tensor
\be
C_{\m\n}=\e_{\m}^{\phantom{\m}\a\b}\na_\a \lt( R_{\b\n}-\f{1}{4}g_{\b\n}R \rt).
\ee

The metric of the warped de Sitter space can be written in the static patch as
\be \label{e8}
ds^2=\f{\ell^2}{3-\n^2} \lt( -(1-r^2)d\t^2+\f{dr^2}{1-r^2}+\f{4\n^2}{3-\n^2} (du+rd\t)^2 \rt),
\ee
with $\n$ being defined as
\be \n \equiv \f{\m\ell}{3} .\ee
Here $(\t,u)\in  R^2$, $r^2<1$ for the patches within the cosmological horizons and $r^2>1$ for the patches between the horizons and the future and past timelike infinity $\ma I^\pm$.

The black hole solution in warped dS$_3$ space has the metric
\bea \label{e2}
&&\f{ds^2}{\ell^2}=\f{2\n}{(3-\n^2)^2}\lt(4\n r +\f{3(1+\n^2)\l}{\n} \rt) dt d\th
   +\f{dr^2}{(3-\n^2)(r_h-r)(r+r_h)}  \nn\\
&& \phantom{\f{ds^2}{\ell^2}=}
    +\f{4\n^2}{(3-\n^2)^2}dt^2 +\f{3(1+\n^2)}{(3-\n^2)^2}
     \lt( r^2+2\l r+\f{3(1+\n^2)\l^2}{4\n^2} -\f{(3-\n^2)r_h^2}{3(1+\n^2)} \rt) d \th^2,
\eea
where $\th \sim \th+2\pi$, and $\l, ~ r_h$ are parameters of the black hole. We focus on the case
\be \label{e6} \n^2<3, \ee
and besides, in order for the above solution to be free of naked closed timelike curves, the parameters must be restricted by
\be \label{e7} \l^2>\f{4\n^2r_h^2}{3(\n^2+1)}. \ee
There are two horizons, namely the black hole horizon $r_b=-r_h$ and the cosmological horizon $r_c=r_h$. The physical observers live in the region between the two horizons.

The conserved charges at $\ma I^+$ corresponding to the $\p_t$ and $\p_\th$ Killing vectors are given by
\bea
&& M_c=Q_{\p_t}=\f{(1+\n^2)\ell\l}{4\n(3-\n^2)G},  \nn\\
&& J_c=Q_{\p_\th}=\f{3(1+\n^2)^2\ell\l^2}{32\n^3(3-\n^2)G}-\f{(5\n^2-3)\ell r_h^2}{24\n(3-\n^2)G}.
\eea
Taking into account the contribution of the topological Chern-Simons term, the cosmological entropy turns out to be
\be\label{Sc}
S_c=\f{\pi\ell}{6\n(3-\n^2)G}[ 3(1+\n^2)\l+(5\n^2-3)r_h ].
\ee
The Hawking temperature and angular velocity of the cosmological horizon $r=r_h$ are
\be
T_c=\f{2\n^2 r_h}{\pi \lt[ 3(1+\n^2)\l+4\n^2 r_h \rt]}, ~~~
\O_c=\f{4\n^2}{ 3(1+\n^2)\l+4\n^2 r_h }.
\ee
For the cosmological horizon the first law of thermodynamics is satisfied
\be \label{e5}
\d M_c=T_c \d S_c  +\O_c \d J_c.
\ee

The holographic dual CFT describing the warped dS$_3$ black hole are proposed in \cite{Anninos:2011vd}.
Just as the case of BTZ and warped AdS$_3$ black holes, the warped dS$_3$ black hole could be obtained by global identification of the warped dS$_3$ along a Killing direction. This discrete identification allows us to read the temperatures
\bea\label{temp}
T_L=\f{3(1+\n^2)\l}{8\pi\n^2}, ~~~ T_R=\f{r_h}{2\pi},
\eea
and the conditions (\ref{e6}) and (\ref{e7}) renders that $T_L>T_R$, viz.
\be \label{e29}  \l > \f{4\n^2r_h}{3(1+\n^2)}. \ee
From the analysis of the asymptotic symmetry, the dual CFT central charges are
\bea \label{e4}
 c_L=\f{4\n\ell}{(3-\n^2)G}, ~~~ c_R=\f{(5\n^2-3)\ell}{\n(3-\n^2)G}.
\eea
Though the dual CFT is an Euclidean CFT, we may still apply the Cardy formula to
count the microscopic entropy, which exactly reproduce the cosmological entropy of the black hole
\be
S_c =\f{\pi^2}{3}(c_L T_L+c_R T_R).
\ee
This gives a strong evidence to support the warped dS/CFT correspondence.

On the other hand, the thermodynamics of the warped dS$_3$ black hole at the black hole horizon $r=-r_h$ is a little subtler. The black hole entropy in the TMG theory is composed of two parts, one coming from the Einstein-Hilbert action and the other coming from the Chern-Simons term \cite{Solodukhin:2005ah,Tachikawa:2006sz}. We write the metric of a general black hole solution in the TMG theory into an Arnowitt-Deser-Misner (ADM) form
\be
ds^2=-N^2 dt^2+g_{rr}dr^2+g_{\th\th} ( d\th+N^\th dt )^2,
\ee
and suppose that $\p_t g_{\m\n}=\p_\th g_{\m\n}=0$. The entropy from the Einstein-Hilbert action is the usual one quarter of the horizon area,
\be
S_{EH}=\f{\ma A}{4G}=\lt.\f{\pi \sr{g_{\th\th}}}{2G}\rt|_{r=r_b}.
\ee
The Chern-Simons term's contribution to the entropy could be \cite{Solodukhin:2005ah,Tachikawa:2006sz}
\be
S_{CS}=\f{\ell}{12\n G}\int_0^{2\pi}\o_{01\th}|_{r=r_b}d\th,
\ee
where $\o_{01\th}$ is the spin connection when we take the vielbein as
\be
e^0=N dt, ~~~ e^1=\sr{g_{rr}}dr, ~~~ e^2=\sr{g_{\th\th}}( d\th+N^\th dt ).
\ee
A routine calculation shows that
\be
\o_{01\th}=\f{g_{\th\th}\p_r N^\th}{2N\sr{g_{rr}}}.
\ee
Then the entropy from the Chern-Simons term is
\be
S_{CS}=\lt.\f{\pi\ell g_{\th\th}\p_r N^\th}{12\n G N \sr{g_{rr}}}\rt|_{r=r_b}.
\ee
Using the metric (\ref{e2}), the total entropy of the warped dS$_3$ black hole at the black hole horizon is calculated as
\be
S_b=S_{EH}+S_{CS}=\f{\pi\ell}{6\n(3-\n^2)G}[ 3(1+\n^2)\l-(5\n^2-3)r_h ],
\ee
and the constraints of the parameters (\ref{e6}), (\ref{e7}) and (\ref{e29}) ensures the positivity of the entropy, viz. $S_b>0$. The Hawking temperature and angular velocity of the black hole horizon $r=-r_h$ are
\be
T_b=\f{2\n^2 r_h}{\pi \lt[ 3(1+\n^2)\l-4\n^2 r_h \rt]}, ~~~
\O_b=\f{4\n^2}{ 3(1+\n^2)\l-4\n^2 r_h }.
\ee
In order to make the first law of thermodynamics be satisfied at the black hole horizon
\be
\d M_b=T_b \d S_b  +\O_b \d J_b,
\ee
we have to redefine the ``mass'' $M_b$ and ``angular momentum'' $J_b$ for the black hole horizon as
\bea
&& M_b=-M_c=-\f{(1+\n^2)\ell\l}{4\n(3-\n^2)G},  \nn\\
&& J_b=-J_c=-\f{3(1+\n^2)^2\ell\l^2}{32\n^3(3-\n^2)G}+\f{(5\n^2-3)\ell r_h^2}{24\n(3-\n^2)G}.
\eea
The situation is somehow different from that for the Kerr-Newman-dS black hole. Here we have the negative ``energy'' $M_b$ at the black hole horizon and the positive ``energy'' $M_c$ at the cosmological horizon. However, for Kerr-Newman-dS black hole the opposite happens, and the ``energy'' should be chosen positive at the black hole horizon and negative at the cosmological horizon \cite{Dehghani:2002np,Ghezelbash:2004af}. In any case, the minus sign of the energy comes from the requirement that the temperature of the horizon be positive.

It is tempting to propose a dual CFT to describe the black hole entropy. Such a CFT could
reside at the black hole horizon or past timelike infinity, just as the cases in Kerr/CFT correspondence.
Note that the black hole entropy is different from the cosmological entropy. As the warped dS$_3$ black hole is always locally diffeomorphic to the warped global dS$_3$ spacetime, the analysis on the asymptotic symmetry group should be the same, which suggests that the central charges should be same. Also the temperature of the CFT should be the same, either from the construction of the black hole or from the study of the hidden conformal symmetry. Then it seems to be impossible to reproduce the black hole entropy from the dual CFT. Nevertheless, notice that the black hole entropy could be related to the central charges and the temperatures with the following relation
\be
S_b=\f{\pi^2}{3}(c_L T_L-c_R T_R),
\ee
which differs from the usual Cardy's formula by a sign before the right sector. We will discuss the above relation in the last section.

\section{Scalar Wave Equation}

In this section we investigate the equation of motion of the scalar perturbation under the warped dS$_3$ black hole background. We solve the scalar equation in two separate regions.

\subsection{In the region $-r_h<r<r_h$}

The equation of the massive scalar $\Phi=e^{-i\o t+ik\th}R(r)$ in the background (\ref{e2}) is
\be \label{e11}
\na_\m \na^\m \Phi=\td \D \Phi=m^2 \phi,
\ee
where we have defined the operator
\be
\td \D=\f{1}{\sr{-g}}\p_\m \sr{-g} g^{\m\n} \p_\n.
\ee
The scalar equation can be calculated explicitly as
\bea \label{e3}
&&\p_r(r_h^2-r^2)\p_r R(r)
+\f{\lt[ \lt( 3(1+\n^2)\l+4\n^2 r_h \rt) \o +4\n^2 k \rt]^2}{32\n^4 r_h (r_h-r)}R(r) \nn\\
&&+\f{\lt[ \lt( 3(1+\n^2)\l - 4\n^2 r_h \rt) \o +4\n^2 k \rt]^2}{32\n^4 r_h (r+r_h)}R(r)
=\lt[ \f{m^2\ell^2}{3-\n^2}+\f{3(1+\n^2)\o^2}{4\n^2} \rt] R(r).
\eea
We solve the scalar equation in the region $-r_h<r<r_h$. With the new coordinate
\be
z=-\f{r+r_h}{r_h-r},
\ee
we have the two independent solutions
\bea
&&R_1=(-z)^\a(1-z)^\b F(a,b,c;z),  \nn\\
&&R_2=(-z)^{-\a}(1-z)^\b F(a-c+1,b-c+1,2-c;z),
\eea
where $F(a,b,c;z)$ is the hypergeometric function and
\bea \label{e35}
&&\a=-i\f{\lt( 3(1+\n^2)\l-4\n^2 r_h \rt)\o +4\n^2 k }{8\n^2 r_h},    \nn\\
&&\b=\f{1}{2}+\sr{\f{1}{4} -\f{m^2\ell^2}{3-\n^2} -\f{3(1+\n^2)\o^2}{4\n^2} },  \nn\\
&&\g=-i \f{\lt( 3(1+\n^2)\l+4\n^2 r_h \rt)\o +4\n^2 k }{8\n^2 r_h},  \nn\\
&&a=\a+\b-\g=\b+i\o,  \nn\\
&&b=\a+\b+\g=\b-i\f{3(1+\n^2)\l\o + 4\n^2 k}{4\n^2 r_h},  \nn\\
&&c=1+2\a.
\eea

For physical solution we impose the ingoing boundary condition at the black hole horizon such that we choose the first solution $R_1$ and discard the second solution $R_2$. When $r$ goes from $-r_h$ to $r_h$, $z$ goes from 0 to $-\inf$. When $|z| \ra \inf$
\be
F(a,b,c;z) \sim \f{\G(c)\G(b-a)}{\G(c-a)\G(b)}(-z)^{-a}+\f{\G(c)\G(a-b)}{\G(c-b)\G(a)}(-z)^{-b},
~~  |\arg(-z)|<\pi,
\ee
and then when $r \ra r_h $ we have
\be
R_1 \sim A_{in} (-z)^\g+  A_{out} (-z)^{-\g},
\ee
with
\be
A_{in}=\f{\G(c)\G(b-a)}{\G(c-a)\G(b)}, ~~~ A_{out}= \f{\G(c)\G(a-b)}{\G(c-b)\G(a)}.
\ee
For physical consideration we should have the outgoing boundary condition at the cosmological horizon \cite{Zhidenko:2003wq}, and thus the first term of the above equation must vanish
\be \label{e36} A_{in}=0.  \ee

\subsection{In the region $r>r_h$}

In the region $r>r_h$, the scalar Eq. (\ref{e3}) can be solved in terms of the new variable
\be
x=\f{r-r_h}{r+r_h},
\ee
and there are two independent solutions
\bea
&&\td R_1=x^{-\g}(1-x)^{\b}F(\td a,\td b,\td c;x),\nn\\
&&\td R_2=x^{\g}(1-x)^{\b}F(\td a-\td c+1,\td b-\td c+1,2-\td c;x),
\eea
with $\a, ~ \b, ~ \g$ defined the same as (\ref{e35}) and
\bea
&&\td a=a=\b+\a-\g=\b+i\o,  \nn\\
&&\td b=\b-\a-\g=\b+i\f{3(1+\n^2)\l\o + 4\n^2 k}{4\n^2 r_h},  \nn\\
&&\td c=1-2\g.
\eea

At the cosmological horizon $r\ra r_h$, we choose the boundary condition that the ingoing mode $\td R_2$ vanishes and the outgoing mode $\td R_1$ coincides with the outgoing mode $R_1$ in the region $r<r_h$ calculated in the last subsection with a proper overall coefficient \cite{Anninos:2010gh}. Outside of the cosmological horizon $r>r_h$, $r$ behaves as the timelike coordinate. As $r\ra\infty$,  $x\ra1,~1-x\ra r^{-1}$, we have the late-time behavior
\be \td R_1 \sim C_1 r^{\b-1}+ C_2 r^{-\b}, \ee
with
\be\label{C21}
C_1=\f{\G(2\b-1)\G(\td c)}{\G(\td a)\G(\td b)},
~~~ C_2=\f{\G(1-2\b)\G(\td c)}{\G(\td c-\td a)\G(\td c-\td b)}.
\ee
We suppose that the scalar solution has the asymptotic behavior $R \sim r^{-h}$ so that we may identify the conformal weights
\bea
&&h_L=h_R=\b=\f{1}{2}+\sr{\f{1}{4} -\f{m^2\ell^2}{3-\n^2} -\f{3(1+\n^2)\o^2}{4\n^2} }, ~~~ \textrm{or} \nn\\
&&h_L=h_R=1-\b=\f{1}{2}-\sr{\f{1}{4} -\f{m^2\ell^2}{3-\n^2} -\f{3(1+\n^2)\o^2}{4\n^2} }.
\eea
The study from the hidden conformal symmetry in Sec. 5 will confirm this identification.

\subsection{Two-point functions}

Before we compute the boundary-boundary correlators, we would like to discuss a subtlety in setting up the warped dS/CFT correspondence. We expect that the quantum gravity asymptotic to warped global dS$_3$ could be holographically described by a conformal field theory. However, the warped dS$_3$ black hole could  have the same asymptotic structure only after a local coordinate transformation.  When $r \ra \inf$ the warped dS$_3$ space metric (\ref{e8}) becomes
\be \label{e30}
ds^2=\f{\ell^2}{3-\n^2} \lt( r^2 d\t^2 -\f{dr^2}{r^2}+\f{4\n^2}{3-\n^2} (du+rd\t)^2 \rt).
\ee
And in the region $r\ra\inf$, the black hole metric (\ref{e2}) behaves as
\be
ds^2=\f{\ell^2}{3-\n^2} \lt( r^2 d\th^2 -\f{dr^2}{r^2}+\f{4\n^2}{3-\n^2} (dt+rd\th)^2 \rt).
\ee
To identify these two metrics, we have to impose the relations
\be
\t=-\th, ~~~ u=-t.
\ee
Note that the above relations only make sense locally as $\th$ is periodic while $\t$ is not. As the translations along $\t$ and $u$ are the Killing symmetry of the warped dS$_3$, we may expand the functions with respect to $\t$ and $u$ as well. This allows us to
define the quantum numbers $\td \o,~\td k$ by the relation
\be
e^{-i\td \o \t+i\td k u}=e^{-i\o t+i k \th}.
\ee
We then have
\be \label{e27}
\td \o=k, ~~~ \td k=\o.
\ee
As the warped dS/CFT correspondence is on the gravity asymptotic to warped dS$_3$, we have to use the quantum numbers $\td \o,~ \td k$ in order to compare the result with that in the CFT side. This is similar to what happens in warped AdS/CFT correspondence\cite{Chen:2009hg}.

Using the quantum numbers $\td \o$ and $\td k$,
the coefficients $\td a, ~ \td b$ in the last subsection can be expressed in terms of the CFT parameters: the conformal weights $(h_L,h_R)$, the frequencies $(\o_L,\o_R)$, the charges $(q_L,q_R)$ and the chemical potentials $(\m_L,\m_R)$
\be
\td a=h_L + i\f{\o_L - q_L \m_L}{2\pi T_L},~~~
\td b=h_R + i\f{\o_R - q_R \m_R}{2\pi T_R},
\ee
and so
\be
i\g=\f{\o_L - q_L \m_L}{4\pi T_L}+\f{\o_R - q_R \m_R}{4\pi T_R},
\ee
with
\bea
&& h_L=h_R=\b \nn\\
&& \o_L=\f{3(1+\n^2)\l\td k}{4\n^2}, ~~~ \o_R=\td \o,  \nn\\
&& q_L=0, ~~~ \m_L=0,  \nn\\
&& q_R=\td k, ~~~ \m_R=-\f{3(1+\n^2)\l}{4\n^2}.
\eea

Formally, we may define the boundary-boundary thermal correlator  from $C_1, ~ C_2$ in (\ref{C21}) as what was done in \cite{Anninos:2010gh}
\bea \label{e32}
&&G_R \sim \f{C_2}{C_1} \pp \f{\G(\td a)\G(\td b)}{\G(\td c-\td a)\G(\td c-\td b)} \nn\\
&&  \phantom{G_R}  \pp \sin \lt(\pi h_L - i \f{\o_L - q_L \m_L}{2T_L} \rt)
                          \sin \lt(\pi h_R - i \f{\o_R - q_R \m_R}{2T_R} \rt)  \nn\\
&&\phantom{G_R\pp} \times \G \lt( h_L-i \f{\o_L - q_L \m_L}{2\pi T_L} \rt)
                            \G \lt( h_L+i \f{\o_L - q_L \m_L}{2\pi T_L} \rt)  \nn\\
&&\phantom{G_R\pp} \times\G \lt( h_R-i \f{\o_R - q_R \m_R}{2\pi T_R} \rt)
                            \G \lt( h_R+i \f{\o_R - q_R \m_R}{2\pi T_R} \rt).
\eea
However, as the future timelike infinity is spacelike, it is not clear for the physical meaning of such boundary-boundary correlator. One possible way to understand it is to do double Wick rotation such that the radial direction becomes really spacelike and the translation along $t$ becomes timelike. Then the above correlator could be considered as a retarded Green's function. Nevertheless, the above correlators is reminiscent of the Euclidean CFT correlators. Recall that in a 2D CFT, by conformal symmetry the Euclidean correlator
takes the form\cite{Cardy:1984bb}
\bea
&&G_E \sim T_L^{2h_L-1}T_R^{2h_R-1}
         e^{ i  (\o_{L,E}-i q_L \m_L)/2T_L }e^{ i  (\o_{R,E}-i q_R \m_R)/2T_R }\nn\\
&&\phantom{G_E \sim}        \times \G \lt( h_L- \f{\o_{L,E}-i q_L \m_L}{2\pi T_L} \rt)
                                   \G \lt( h_L+ \f{\o_{L,E}-i q_L \m_L}{2\pi T_L} \rt) \nn\\
&&\phantom{G_E \sim}        \times \G \lt( h_R- \f{\o_{R,E}-i q_R \m_R}{2\pi T_R} \rt)
                                   \G \lt( h_R+ \f{\o_{R,E}-i q_R \m_R}{2\pi T_R} \rt),
\eea
with the Euclidean frequencies
\be
\o_{L,E}=i\o_L, ~~~
\o_{R,E}=i\o_R.
\ee

As the conformal weight could be smaller than 1, we may alternate the roles of the source and the response and define the boundary-boundary thermal correlator as
\be
G_R \sim \f{C_1}{C_2},
\ee
and then we can get the same results as before except that
\be
h_L=h_R=1-\b.
\ee
In both cases, we can read the poles in the correlator (\ref{e32}) as
\be\label{poles}
\o_R=q_R\m_R-i2\pi T_R (n+h_R).
\ee

From the first law of black hole thermodynamics (\ref{e5}) and its CFT counter-term
\be
\d S_{CFT}=\f{\d E_L}{T_L}+\f{\d E_R}{T_R},
\ee
we get the conjugate charges $(\d E_L,\d E_R)$ as
\bea
&&\d E_L=\f{3(1+\n^2)\l}{4\n^2}\d M_c, \nn\\
&&\d E_R=\f{3(1+\n^2)\l}{4\n^2}\d M_c-\d J_c.
\eea
The identifications of parameters are $\d M_c=\o=\td k,~\d J_c=-k=-\td \o$, and then we have
\bea
&&\o_L=\d E_L \lt( \d M_c=\td k;\d J_c=-\td \o \rt), \nn\\
&&\o_R - q_R \m_R=\d E_R \lt( \d M_c=\td k;\d J_c=-\td \o \rt).
\eea
Here the unusual minus sign comes from the definitions of the horizon angular velocity and angular momentum in \cite{Anninos:2011vd}.

\section{Quasinormal Modes}

In this section we calculate the quasinormal modes of scalar, vector, tensor and spinor perturbations under the warped dS$_3$ black hole background.

\subsection{Scalar perturbation}

The constraint (\ref{e36}) requires
\be
b=-n, ~~ \textrm{or} ~~ c-a=-n,
\ee
with $n$ being a nonnegative integer. When $b=-n$, we have
\be \label{e20}
\o=-\f{k}{2\pi T_L}-i\f{T_R}{T_L}(n+\b),
\ee
and when $c-a=-n$, we have
\be \label{e21}
\o=-\f{k}{2\pi T_L}-i\f{T_R}{T_L}(n+1-\b).
\ee

Taking into account of the reidentification of the quantum numbers (\ref{e27}),
then $b=-n$ gives us the quasinormal modes of the frequencies
\bea \label{e9}
&&\td \o_R=-2\pi T_L \td k-i 2\pi T_R (n+h_R^+), \nn\\
&&h_R^+=\b=\f{1}{2}+\sr{\f{1}{4} -\f{m^2\ell^2}{3-\n^2} -\f{3(1+\n^2)\td k^2}{4\n^2} },
\eea
and $c-a=-n$ gives us the quasinormal modes of the frequencies
\bea \label{e10}
&&\td \o_R=-2\pi T_L \td k-i 2\pi T_R (n+h_R^-), \nn\\
&&h_R^-=1-\b=\f{1}{2}-\sr{\f{1}{4} -\f{m^2\ell^2}{3-\n^2} -\f{3(1+\n^2)\td k^2}{4\n^2} },
\eea
We can see that the frequency always has a negative imaginary part and so has the right behavior of the quasinormal modes. Also we can see that the quasinormal modes frequencies (\ref{e9}) and (\ref{e10}) are the poles (\ref{poles}) of the boundary-boundary two-point function.

The warped AdS$_3$ space has the asymptotic behavior \cite{Anninos:2008fx,Chen:2009hg}
\be \label{e31}
ds^2=\f{\ell^2}{\n^2+3}\lt( -r^2d\t^2+\f{dr^2}{r^2}+\f{4\n^2}{\n^2+3}(dx+rd\t)^2 \rt).
\ee
We can see that if we set $\ell \ra i\ell$ (consequently $\n \ra i\n$), the metric (\ref{e30}) becomes exactly (\ref{e31}), so that it is not surprising that the conformal weights of the scalar field of the warped dS$_3$ black hole (\ref{e9}) and (\ref{e10}) become exactly those of the spacelike stretched warped AdS$_3$ black hole \cite{Chen:2009hg,Chen:2010ik}
\be
h_R^\pm=\f{1}{2} \pm \sr{\f{1}{4} +\f{m^2\ell^2}{\n^2+3} -\f{3(\n^2-1)\td k^2}{4\n^2}}.
\ee
Moreover if we set $\ell \ra i\ell$ and $\n\ra i$, the conformal weights become exactly that of the BTZ black hole \cite{Birmingham:2001pj,Chen:2010ik}
\be h_L^\pm=h_R^\pm=\f{1}{2}(1 \pm \sr{1+m^2\ell^2}). \ee

\subsection{Vector perturbation}

 From the vector equation
\be
\e_\l^{\phantom{\l}\m\n}\p_\m A_\n=-m A_\l,
\ee
we have
\be
\td \D A_t= \lt( m^2+\f{2\n m}{\ell} \rt) A_t.
\ee
Comparing the vector equation with the scalar Eq. (\ref{e11}), we can see clearly that in order to get the quasinormal modes of the vector perturbation we just need to make the change $m^2 \ra m^2+\f{2\n m}{\ell}$ in the scalar results. Then we get the quasinormal modes and conformal weights of the vector field
\bea \label{e12}
&&\td \o_R=-2\pi T_L \td k-i 2\pi T_R (n+h_R^\pm), \nn\\
&&h_R^\pm=\f{1}{2}\pm\sr{\f{1}{4} - \f{m^2\ell^2 +2\n m\ell}{3-\n^2} -\f{3(1+\n^2)\td k^2}{4\n^2} }.
\eea

There is another vector equation
\be
\e_\l^{\phantom{\l}\m\n}\p_\m A_\n=m A_\l,
\ee
which corresponds to vector field with a different helicity. With this vector equation, we get the same quasinormal modes and slightly different conformal weights
\bea \label{e13}
&&\td \o_R=-2\pi T_L \td k-i 2\pi T_R (n+h_R^\pm), \nn\\
&&h_R^\pm=\f{1}{2}\pm\sr{\f{1}{4} - \f{m^2\ell^2 -2\n m\ell}{3-\n^2} -\f{3(1+\n^2)\td k^2}{4\n^2} }.
\eea

If we set $\ell \ra i\ell$ (consequently $\n \ra i\n$), the conformal weights of the vector field of the warped dS$_3$ black hole (\ref{e12}) and (\ref{e13}) become exactly those of the spacelike stretched warped AdS$_3$ black hole \cite{Chen:2009hg,Chen:2010ik}. Furthermore if we set $\ell\ra i\ell$ and $\n \ra i$, the conformal weights become exactly those of the BTZ black hole \cite{Birmingham:2001pj,Chen:2010ik}
\be h_L=\f{m\ell}{2}, ~~~ h_R=1+\f{m\ell}{2}, \ee
or
\be h_L=1-\f{m\ell}{2}, ~~~ h_R=-\f{m\ell}{2}. \ee
In this case we have $|h_L-h_R|=s$ with $s=1$.

\subsection{Tensor perturbation}

From the equation for tensor perturbation
\be
\e_\l^{\phantom{\l}\m\n} \na_\m h_{\n\s} =\mp m h_{\l\s},
\ee
we have
\be
\td \D h_{tt}= \lt( m^2 \pm \f{4\n m}{\ell}+\f{3\n^2}{\ell^2} \rt) h_{tt}.
\ee
To get the quasinormal modes of the tensor perturbation we just make the change $m^2 \ra m^2 \pm \f{4\n m}{\ell}+\f{3\n^2}{\ell^2}$ in the scalar results, and we get
\bea \label{e14}
&&\td \o_R=-2\pi T_L \td k-i 2\pi T_R (n+h_R^\pm), \nn\\
&&h_R^\pm=\f{1}{2} \pm \sr{\f{1}{4} - \f{m^2\ell^2 + 4\n m\ell+3\n^2}{3-\n^2} -\f{3(1+\n^2)\td k^2}{4\n^2} },
~~ \textrm{or} \nn\\
&&h_R^\pm=\f{1}{2} \pm \sr{\f{1}{4} - \f{m^2\ell^2 - 4\n m\ell+3\n^2}{3-\n^2} -\f{3(1+\n^2)\td k^2}{4\n^2} }.
\eea

If we set $\ell \ra i\ell$ (consequently $\n \ra i\n$), the conformal weights of the tensor field of the warped dS$_3$ black hole (\ref{e14}) become that of the spacelike stretched warped AdS$_3$ black hole \cite{Chen:2009hg,Chen:2010ik}. Furthermore if we set $\ell \ra i\ell$ and $\n \ra i$, the conformal weights become exactly those of the BTZ black hole \cite{Chen:2010ik}
\bea
&&h_L=\f{m\ell-1}{2}, ~~~ h_R=\f{m\ell+3}{2},
~~ \textrm{or} \nn\\
&&h_L=\f{-m\ell+3}{2}, ~~~ h_R=\f{-m\ell-1}{2}.
\eea
In this case we have $|h_L-h_R|=s$ with $s=2$.

\subsection{Spinor perturbation}

In order to solve the Dirac equations in the warped dS$_3$ black hole, we should choose the vielbein for the background and calculate the corresponding spin connection. To simplify the process we parameterize the metric of warped $dS_3$ black hole in the form
\bea
ds^2=g_{tt}dt^2+\f{dr^2}{g^{rr}}+2g_{t\th}dtd\th+\f{g g^{rr}+g_{t\th}^2}{g_{tt}}d\th^2,
\eea
where we have two functions
\be
g^{rr}=\f{(r+r_h)(r_h-r)}{-A_1}, ~~~ g_{t\th}=A_2 r+A_3,
\ee
and five constants
\bea
&&g_{tt}=\f{4\n^2\ell^2}{(3-\n^2)^2}, ~~~ g=-\f{4\n^2\ell^6}{(3-\n^2)^4},  \nn\\
&& A_1=-\f{\ell^2}{3-\n^2}, ~~~ A_2=\f{4\n^2\ell^2}{(3-\n^2)^2},
~~~ A_3=\f{3(1+\n^2)\l\ell^2}{(3-\n^2)^2}.
\eea
The vielbeins $e_\m^a$ are chosen as
\be
e^0=\sr{\f{-g g^{rr}}{g_{tt}}}d\th, ~~~ e^1=\f{dr}{\sr{g^{rr}}},
 ~~~ e^2=\sr{g_{tt}}dt+\f{g_{t\phi}}{\sr{g_{tt}}}d\th,
\ee
where $e^a=e^a_\m dx^\m$. The spin connection can be calculated straightforwardly and the nonvanishing components of the spin connection are
\bea
&& \o_t^{01}=-\o_t^{10}=-\f{g'_{t\th}}{2}\sr{\f{g_{tt}}{-g}}, ~~~
\o_\th^{01}=-\o_\th^{10}=-\f{g g'^{rr}+g_{t\th}g'_{t\th}}{2\sr{-g g_{tt}}}, \nn\\
&& \o^{02}_r=-\o_r^{20}=-\f{g'_{t\th}}{2\sr{-g g^{rr}}}, ~~~
\o^{12}_\th=-\o_\th^{21}=-\f{g'_{t\th}}{2}\sr{\f{g^{rr}}{g_{tt}}}.
\eea

The Dirac equation is
\be
\g^a e^\m_a \lt( \p_\m +\f{1}{2}\o_\m^{ab}\S_{ab} \rt)\Psi + m\Psi=0,
\ee
where $\S_{ab}=\f{1}{4}[\g_a,\g_b], ~ \g^0=i\s^2, ~ \g^1=\s^1, ~ \g^2=\s^3$. With the ansatz $\Psi=(\psi_+,\psi_-)e^{-i\o t+ik\th}$, we have
\bea
&& \sr{g^{rr}}\p_r \psi_++\f{g'^{rr}}{4\sr{g^{rr}}}\psi_+
   -\f{i(g_{t\th}\o+g_{tt}k)}{\sr{-g g_{tt} g^{rr}}}\psi_+
   +\lt( m-\f{g'_{t\th}}{4\sr{-g}} +\f{i\o}{\sr{g_{tt}}} \rt)\psi_-=0, \nn\\
&& \sr{g^{rr}}\p_r \psi_-+\f{g'^{rr}}{4\sr{g^{rr}}}\psi_-
   +\f{i(g_{t\th}\o+g_{tt}k)}{\sr{-g g_{tt} g^{rr}}}\psi_-
   +\lt( m-\f{g'_{t\th}}{4\sr{-g}} -\f{i\o}{\sr{g_{tt}}} \rt)\psi_+=0.
\eea
Then the equations can be transformed into the form
\bea
&& \p_r(r_h^2-r^2)\p_r \psi_+
+\f{\lt[ 4A_1\lt( (A_3-A_2 r_h)\o+g_{tt} k \rt) -i\sr{-g g_{tt}}2r_h \rt]^2}
   {-32gg_{tt}r_h(r+r_h)}\psi_+    \nn\\
&&+\f{\lt[ 4A_1\lt( (A_3+A_2 r_h)\o+g_{tt} k \rt) +i\sr{-g g_{tt}}2r_h \rt]^2}
     {-32gg_{tt}r_h(r_h-r)}\psi_+    \nn\\
&&=\lt( \f{1}{4} -A_1(m-\f{A_2}{4\sr{-g}})^2 +\f{A_1(A_1 A_2^2+g)\o^2}{-g g_{tt}} \rt)\psi_+, \nn\\
\eea
\bea
&& \p_r(r_h^2-r^2)\p_r \psi_-
+\f{\lt[ 4A_1\lt( (A_3-A_2 r_h)\o+g_{tt} k \rt) +i\sr{-g g_{tt}}2r_h \rt]^2}
   {-32gg_{tt}r_h(r+r_h)}\psi_-    \nn\\
&&+\f{\lt[ 4A_1\lt( (A_3+A_2 r_h)\o+g_{tt} k \rt) -i\sr{-g g_{tt}}2r_h \rt]^2}
     {-32gg_{tt}r_h(r_h-r)}\psi_-    \nn\\
&&=\lt( \f{1}{4} -A_1(m-\f{A_2}{4\sr{-g}})^2 +\f{A_1(A_1 A_2^2+g)\o^2}{-g g_{tt}} \rt)\psi_-. \nn\\
\eea

The solutions of these equations with only the ingoing flux at the black hole event horizon $r=-r_h$ are given by the hypergeometric functions
\bea
&& \psi_+ \pp z^{\a_f}(1-z)^{\b_f}F(a_f,b_f,c_f;z) , \nn\\
&& \psi_- \pp z^{\a_f-1/2}(1-z)^{\b_f}F(a_f-1,b_f,c_f-1;z),
\eea
with the new coordinate $z$ being
\be
z=-\f{r+r_h}{r_h-r},
\ee
and the parameters
\bea
&&\a_f=-i\f{\lt( 3(1+\n^2)\l-4\n^2 r_h \rt)\o +4\n^2 k }{8\n^2 r_h} +\f{1}{4},    \nn\\
&&\b_f=\f{1}{2}+\sr{ -\f{(m\ell-\n/2)^2}{3-\n^2} -\f{3(1+\n^2)\o^2}{4\n^2} },  \nn\\
&&\g_f=-i \f{\lt( 3(1+\n^2)\l+4\n^2 r_h \rt)\o +4\n^2 k }{8\n^2 r_h} -\f{1}{4},  \nn\\
&&a_f=\a_f+\b_f-\g_f=\b_f+\f{1}{2}+i\o,  \nn\\
&&b_f=\a_f+\b_f+\g_f=\b_f-i\f{3(1+\n^2)\l\o + 4\n^2 k}{4\n^2 r_h},  \nn\\
&&c_f=1+2\a_f.
\eea

Just like the scalar case, we impose the outgoing boundary condition at the cosmological horizon $r=r_h$, and get the constraints
\be
b_f=-n, ~~ \textrm{or} ~~ c_f-a_f=-n.
\ee
With the above constraints and the identifications of the quantum numbers (\ref{e27}), we have the quasinormal modes of the fermion perturbation in the warped dS$_3$ black hole
\bea \label{e28}
&&\td \o_R=-2\pi T_L \td k-i 2\pi T_R (n+h_R^\pm), \nn\\
&&h_R^\pm=\f{1}{2}\pm\sr{-\f{(m\ell-\n/2)^2}{3-\n^2} -\f{3(1+\n^2)\td k^2}{4\n^2}}.
\eea
Note that the conformal weights are always complex for fermionic perturbations.

If we set $\ell \ra i\ell$ (consequently $\n \ra i\n$), the conformal weights of the vector field of the warped dS$_3$ black hole (\ref{e28}) become those of the spacelike stretched warped AdS$_3$ black hole \cite{Chen:2009hg}. Furthermore if we set $\ell\ra i\ell$ and $\n \ra i$, the conformal weights (\ref{e28}) become those of the BTZ black hole \cite{Birmingham:2001pj}
\be h_L=\f{1}{4}+\f{m\ell}{2}, ~~ \textrm{or} ~~  h_L=\f{3}{4}-\f{m\ell}{2}. \ee

\section{Quasinormal Modes from the Hidden Conformal Symmetry}

Because of the existence of the cosmological horizon, the hidden conformal symmetry of the de Sitter black holes is a little subtler. In the region outside of the cosmological horizon $r>r_c$, we can analyze the hidden conformal symmetry as the ordinary black hole; however, here we just do the analysis in the region between the black hole horizon and the cosmological horizon $r_b<r<r_c$. We also construct the quasinormal modes of the warped dS$_3$ black hole from the hidden conformal symmetry.

\subsection{Hidden conformal symmetry}

In the region $r_b<r<r_c$, we define the conformal coordinates
\bea
&&\o^+=\sr{\f{r_c-r}{r-r_b}}e^{2\pi T_R\th+2n_R t},  \nn\\
&&\o^-=\sr{\f{r_c-r}{r-r_b}}e^{2\pi T_L\th+2n_L t},  \nn\\
&&y=\sr{\f{r_c-r_b}{r-r_b}}e^{\pi(T_R+T_L)\th+(n_R+n_L )t},
\eea
with which the vector fields $(V_0,V_\pm)$ and $(\td V_0, \td V_\pm)$ could be locally defined as
\bea
&&V_1=\p_+, \nn\\
&&V_0=\o^+\p_++\frac{1}{2}y\p_y, \nn\\
&&V_{-1}=\o^{+2}\p_++\o^+y\p_y+y^2\p_-,
\eea
and
\bea
&&\tilde V_1=\p_- \nn\\
&&\tilde V_0=\o^-\p_-+\frac{1}{2}y\p_y \nn\\
&&\tilde V_{-1}=\o^{-2}\p_-+\o^-y\p_y+y^2\p_+.
\eea
These vector fields obey the $SL(2,R)$ Lie algebra
\be
[V_0, V_{\pm 1}]=\mp V_{\pm 1},\hs{5ex} [V_{-1},V_1]=-2 V_0,
\ee
and similarly for $(\tilde V_0, \tilde V_{\pm 1})$. Written in the coordinates $(t,r,\th)$, these vector fields are
\bea \label{e22}
&&V_1=e^{-2\pi T_R \th-2 n_R t} \lt( -\sr{\D}\p_r -\f{T_L \D'-T_R(r_c-r_b)}{4A\sr{\D}}\p_t
                                     +\f{n_L \D'-n_R(r_c-r_b)}{4 \pi A\sr{\D}}\p_\th \rt),  \nn\\
&&V_0=\f{\pi T_L \p_t-n_L\p_\th}{2\pi A}, \nn\\
&&V_{-1}=e^{2\pi T_R \th+2 n_R t} \lt( -\sr{\D}\p_r +\f{T_L \D'-T_R(r_c-r_b)}{4A\sr{\D}}\p_t
                                     -\f{n_L \D'-n_R(r_c-r_b)}{4 \pi A\sr{\D}}\p_\th \rt),  \nn\\
\eea
and
\bea \label{e23}
&&\td V_1=e^{-2\pi T_L \th-2 n_L t} \lt( -\sr{\D}\p_r +\f{T_R \D'-T_L(r_c-r_b)}{4A\sr{\D}}\p_t
                                     -\f{n_R \D'-n_L(r_c-r_b)}{4 \pi A\sr{\D}}\p_\th \rt),  \nn\\
&&\td V_0=\f{-\pi T_R \p_t+n_R\p_\th}{2\pi A}, \nn\\
&&\td V_{-1}=e^{2\pi T_L \th+2 n_L t} \lt( -\sr{\D}\p_r -\f{T_R \D'-T_L(r_c-r_b)}{4A\sr{\D}}\p_t
                                     +\f{n_R \D'-n_L(r_c-r_b)}{4 \pi A\sr{\D}}\p_\th \rt),  \nn\\
\eea
where we have defined  \be \D=(r_c-r)(r-r_b), ~~~ A=n_R T_L - n_L T_R.  \ee

The quadratic Casimir is defined as
\bea
&&\ma H^2=\tilde{\ma H}^2=-V_0^2+\frac{1}{2}(V_1 V_{-1}+V_{-1}V_1) \nn\\
&&\phantom{\ma H^2}=\frac{1}{4}( y\p_y - y^2\p^2_y) +y^2 \p_+\p_-.
\eea
and in terms of $(t,r,\th)$ coordinates it becomes
\bea
&&\ma H^2=\p_r \D \p_r
-\f{(r_c-r_b)[\pi(T_L+T_R)\p_t-(n_L+n_R)\p_\th]^2}{16\pi^2 A^2 (r_c-r)} \nn\\
&&\phantom{\ma H^2=} -\f{(r_c-r_b)[\pi(T_L-T_R)\p_t-(n_L-n_R)\p_\th]^2}{16\pi^2 A^2 (r-r_b)}.
\eea
With the scalar field being expanded as $\Phi=e^{-i\o t+ik\th}R(r)$, the equation $\ma H^2\Phi=-K\Phi$  gives us the radial equation of motion
\bea \label{e18}
&&\p_r \D \p_r R(r)
+\f{(r_c-r_b)[\pi(T_L+T_R)\o+(n_L+n_R)k]^2}{16\pi^2 A^2 (r_c-r)} R(r) \nn\\
&&+\f{(r_c-r_b)[\pi(T_L-T_R)\o+(n_L-n_R)k]^2}{16\pi^2 A^2 (r-r_b)} R(r)=-K R(r),
\eea
where $K$ is a constant.

The Eq. (\ref{e18}) is the same as the Eq. (\ref{e3}) under the identifications
\bea
&&r_c=r_h, ~~~ r_b=-r_h, \nn\\
&&K=-\f{m^2\ell^2}{3-\n^2}-\f{3(1+\n^2)\o^2}{4\n^2},  \nn\\
&&T_L=\f{3(1+\n^2)\l}{8\pi\n^2}, ~~~ n_L=\f{1}{2}, \nn\\
&&T_R=\f{r_h}{2\pi}, ~~~ n_R=0.
\eea
This suggests that the warped dS$_3$ black hole does have the hidden conformal symmetry in the whole physical region. Another remarkable point is that in the above identifications we find the same dual temperatures $T_L$ and $T_R$ as the ones (\ref{temp}) suggested in \cite{Anninos:2011vd}. This provides another support that the black hole could be described by a finite temperature CFT.

\subsection{Quasinormal modes construction}

In this subsection we construct the quasinormal modes of the warped dS$_3$ black hole from the hidden conformal symmetry in an algebraic way based on the formalism proposed in \cite{Chen:2010ik,Chen:2010sn}, where the readers can see more details. We use the hidden conformal symmetry in the physical region and thus the vector fields (\ref{e22}) and (\ref{e23}).

Under the background of a warped black hole, the equation of motion of scalar, and some linear combinations of the vector or tensor components $\Phi$ can be written in the form
\be
\lt( \ma L^2 +K \rt) \Phi=\lt( \ma L^2 +b\ma L^2_{\td V_0} +a \rt) \Phi=0,
\ee
where $\ma L$ denotes the Lie-derivative and $\ma L^2$ denotes the Lie-induced quadratic Casimir
\be
\ma L^2 \equiv -\ma L_{V_0} \ma L_{V_0} +\f{1}{2} \lt( \ma L_{V_1}\ma L_{V_{-1}}
                                                       + \ma L_{V_{-1}}\ma L_{V_1} \rt),
\ee
and $K$, $a$ and $b$ are some constants to be fixed.

As the construction of various kinds of quasinormal modes is similar, here we take the scalar field as the example to illustrate the method. Firstly, we define the highest weight state as
\be \ma L_{V_0}\Phi^{(0)}=h_R \Phi^{(0)}, ~~~ \ma L_{V_1}\Phi^{(0)}=0, \ee
and then, from the highest weight state we construct the descendent modes
\be \Phi^{(n)}=\ma L^n_{V_{-1}}\Phi^{(0)}. \ee
We also define $\ma L_{\td V_0}\Phi=q\Phi$, then we have
\be K=b q^2+a. \ee
Since the Lie-induced Casimir $\ma L^2$ always commute with the Lie-derivatives, with $\ma L^2$ operating on the descendent modes $\Phi^{(n)}$, we have
\be h_R^2-h_R-K=0, \ee
and then we get the conformal weight
\be
h_R=\f{1}{2} \pm \sr{\f{1}{4}+K}.
\ee
To compute the frequency of the quasinormal modes, we expand $\Phi=e^{-i\o t+ik\th}$. With $\ma L_{V_0}$ operating on the descendent modes $\Phi^{(n)}$ we get the quasinormal modes
\be \label{e24}
\o=-\f{n_L}{\pi T_L}k+i\f{2A}{T_L}(n+h_R).
\ee
Note that only when $A<0$, the above frequency have the right behavior of quasinormal modes. 

When $A<0$, we can solve the highest condition $\ma L_{V_1}\Phi^{(0)}=0$ as
\be
R^{(0)}=C (r_c-r)^{i\f{\pi(T_L+T_R)\o+(n_L+n_R)k}{4\pi A}}
          (r-r_b)^{i\f{\pi(T_L-T_R)\o+(n_L-n_R)k}{4\pi A}},
\ee
where $C$ is a constant. When $r \ra r_b+0$, we have
\be
\Phi^{(0)} \sim e^{-i\o \lt( t-\f{\pi(T_L-T_R)}{4\pi A}\ln(r-r_b) \rt)},
\ee
which just has the ingoing boundary condition. Also when $r \ra r_c-0$, we have
\be
\Phi^{(0)} \sim e^{-i\o \lt( t-\f{\pi(T_L+T_R)}{4\pi A}\ln(r_c-r) \rt)},
\ee
which just has the outgoing boundary condition. We can see $\Phi^{(0)}$, and thus all $\Phi^{(n)}$ have the right behavior as the quasinormal modes. 

In the case of warped dS$_3$ black hole, we have
\bea
&&b=\f{3(1+\n^2)}{4\n^2}, ~~~ q=-i\o, ~~~ a=-\f{m^2\ell^2}{3-\n^2},  \nn\\
&&K=-\f{m^2\ell^2}{3-\n^2}-\f{3(1+\n^2)\o^2}{4\n^2}.
\eea
After the reidentification of the quantum numbers (\ref{e27}) we get the same quasinormal modes for the scalar perturbation (\ref{e9}) and (\ref{e10}). For the vector and tensor perturbations the quasinormal modes can be constructed similarly, and the results all agree with that of those Sec. 3. Unlike $\td V_0$, the vector $V_0$ has a nonvanishing $\th$ component, so we can not use another set of vectors $(\td V_0, \td V_{\pm1})$ to construct the quasinormal modes for the warped dS$_3$ black hole.

\section{Conclusion and Discussion}

In this paper we studied several aspects  of the warped dS/CFT correspondence. We analyzed the scalar wave function carefully and got the boundary-boundary correlators at the future timelike infinity. After imposing the ingoing boundary conditions at the black hole event horizon and outgoing boundary conditions at the cosmological horizon, we read the quasinormal modes of various perturbations of the black hole. The frequencies of the quasinormal modes of various kinds of perturbations all take the form
\be
\td \o_R=-2\pi T_L \td k-i 2\pi T_R (n+h_R), \nn\\
\ee
 which correspond to the poles in the boundary-boundary correlators. Moreover we investigated the hidden conformal symmetry of the warped dS$_3$ black hole in the physical region and constructed the towers of the quasinormal modes with the help of the ladder operators, which are in perfect match with what were obtained by solving the equations of motion.

One interesting point in the dS/CFT correspondence is the existence of the complex conformal weight. It is easy to see that the conformal weights of scalar perturbation (\ref{e9}) and (\ref{e10}), vector perturbations (\ref{e12}) and (\ref{e13}), and tensor perturbations (\ref{e14}) are all complex for large enough mass, similar to the cases in other dS/CFT situations \cite{Strominger:2001pn} and \cite{Anninos:2009yc}. For spinor perturbations the conformal weights (\ref{e28}) are always complex. The complexity of the conformal weight may indicate that the theory is nonunitary. Another novel feature in the warped
dS/CFT correspondence is that the conformal weights depend on the quantum number, besides the mass and spin, similar to the warped AdS/CFT correspondence \cite{Chen:2009hg}.

One puzzle on the warped dS$_3$ black hole is how to interpret its black hole entropy. Unlike its cousin in dS$_3$,  there are two horizons for a warped dS$_3$ black hole. In Sec. 2, we computed its entropy, which could be rewritten in a Cardy-like formula but with an opposite sign before the contribution from the right sector.
Without justification, we suppose that for the warped dS$_3$ black hole there is also CFT dual at the black hole horizon or at the past timelike infinity. In order to write the entropy of the black hole horizon in form of the Cardy formula
\be
S_b=\f{\pi^2}{3}(c_L^b T_L^b + c_R^b T_R^b),
\ee
we need
\bea
&& c_L^b=\f{4\n\ell}{(3-\n^2)G}, ~~~ c_R^b=\f{(3-5\n^2)\ell}{\n(3-\n^2)G}, \nn\\
&& T_L^b=\f{3(1+\n^2)\l}{8\pi\n^2}, ~~~ T_R^b=\f{r_h}{2\pi}.
\eea
Note that the right moving central charges of the CFT duals at the two horizons are of the opposite sign. On the other hand,  for a given Virasoro algebra
\be
[L_m,L_n]=(m-n)L_{m+n}+\f{c}{12}(m^3-m)\d_{m+n},
\ee
we can always redefine the generators $\hat L_m=-L_{-m}$ and obtain a Virasoro algebra with opposite central charge
\be
[\hat L_m,\hat L_n]=(m-n)\hat L_{m+n}+\f{-c}{12}(m^3-m)\d_{m+n}.
\ee
In other words, from the analysis of asymptotic symmetry group, it is impossible to fix the right central charge. The only criterion is that the central charge should be positive. In the case at hand, we may have different central charges by going to the different parameter region. Therefore, for one possible value $\n^2<3$, we can always have a well-defined central charge, suitable for describing either the black hole entropy or cosmological entropy, but not both. This is a puzzle we find no satisfying way to resolve. One possible way is that the theory is not well-defined in general. Only when $\n^2=\f{3}{5}$, the right central charge is vanishing, the two CFTs could accommodate with each other. This is reminiscent of the chiral gravity in AdS$_3$ \cite{chiral}. It would be very interesting to check if such a chiral point make sense in the warped dS$_3$.

\vspace*{10mm}

\noindent
 {\large{\bf Acknowledgments}}

We thank Jia-Rui Sun for valuable discussions. The work was in part supported by NSFC Grant No. 10975005.

\end{document}